\documentstyle[11pt,aaspp4]{article}
\def\puncspace{\ifmmode\,\else{\ifcat.\C{\if.\C\else\if,\C\else\if?\C\else%
\if:\C\else\if;\C\else\if-\C\else\if)\C\else\if/\C\else\if]\C\else\if'\C%
\else\space\fi\fi\fi\fi\fi\fi\fi\fi\fi\fi}%
\else\if\empty\C\else\if\space\C\else\space\fi\fi\fi}\fi}
\def\SP{\let\\=\empty\futurelet\C\puncspace }
\def\etal{et\SP al.\SP }

\def\kms{kms$^{-1}$}
\def\h-1{$h^{-1}$}
\def\void#1{{}}

\def\h1{$h^{-1}$}

\def\iras{$IRAS$\SP}

\def\kms{kms$^{-1}$ }
\def\etal{et al.\,}
\def\eg{e.g., \,}

\def\xip {$\xi(r_p,\pi)\>$}
\def\wrp {$\omega(r_p)\>$\SP} 
\def\sig8 {$\sigma_8\>$} 
\def\xis {$\xi (s)\>$\SP}
\def\xir {$\xi (r)\>$}
\def\ap {$\sim\>$}
\def\mm{$\pm$}
\def\lsim{~\rlap{$<$}{\lower 1.0ex\hbox{$\sim$}}}
\def\gsim{~\rlap{$>$}{\lower 1.0ex\hbox{$\sim$}}}
\begin{document}

\title{Southern Sky Redshift Survey: Clustering of Local Galaxies \footnote{Based on observations at
Cerro Tololo Interamerican Observatory (CTIO), National Optical
Astronomy Observatories (NOAO) which is operated by the Association of
Universities for Research in Astronomy, Inc. under contract to the
National Science Foundation; 
Complejo Astronomico El Leoncito (CASLEO), operated under agreement between the
Consejo Nacional de Investigaciones Cient\'\i ficas de la Rep\'ublica
Argentina and the National Universities of La Plata, C\'ordoba and San
Juan; European Southern Observatory (ESO), partially under the ESO-ON
agreement; Laborat\'orio Nacional de Astrof\'\i sica; and the South
African Astronomical Observatory}}

\author{C. N. A. Willmer}
\affil{Observat\'orio Nacional, Rua General Jos\'e Cristino
77, Rio de Janeiro, RJ 20921-030, Brazil} 
\affil{electronic mail: cnaw@on.br}
\author{L. Nicolaci da Costa \altaffilmark{2}}
\affil{European Southern Observatory, Karl-Schwarzchild Str. 2,
Garching-bei-M\"unchen, Germany}
\affil{electronic mail: ldacosta@eso.org}
\altaffiltext{2}{Observat\'orio Nacional, Rua General Jos\'e Cristino
77, Rio de Janeiro, RJ 20921-030, Brazil.}
\and

\author{P. S. Pellegrini}
\affil{Observat\'orio Nacional, Rua General Jos\'e Cristino
77, Rio de Janeiro, RJ 20921-030, Brazil} 
\affil{electronic mail: pssp@on.br}

\begin{abstract} We use the two-point correlation function to calculate
the clustering properties of the recently completed SSRS2 survey, which
probes two well separated regions of the sky, allowing one to evaluate
the sensitivity of sample-to-sample variations. Taking advantage
of the large number of galaxies in the combined sample, we also
investigate the dependence of clustering on the internal properties of
galaxies.

The redshift space correlation function
for the combined magnitude-limited sample of the SSRS2 is given by 
$\xi(s)=(s/5.85$ \h1 Mpc)$^{-1.60}$ for separations between 2 $\leq s
\leq$ 11 \h1 Mpc, while our best estimate for the real space correlation
function is $\xi (r) = (r/5.36$ \h1 Mpc)$^{-1.86}$. Both are comparable
to previous measurements using surveys of optical galaxies over much
larger and independent volumes. By comparing the correlation function
calculated in redshift and real space we find that the redshift
distortion on intermediate scales is small.
This result implies that the observed redshift-space
distribution of galaxies is close to that in  real space, and that
$\beta = \Omega^{0.6}/b < 1$, where $\Omega$ is the cosmological density
parameter and  $b$ is the linear biasing factor for optical galaxies.

We have used the SSRS2 sample to study the dependence of $\xi$ on the
internal properties of galaxies such as luminosity, morphology and
color.  We confirm earlier results that luminous galaxies ($L>L^*$) are
more clustered than sub-$L^*$ galaxies and that the luminosity
segregation is scale-independent. We also find that early types are
more clustered than late types. However, in the absence of rich clusters,
the relative bias between early and late types in real space,
$b_{E+S0}$/$b_S$ $\sim$ 1.2, is not as strong 
as previously estimated. Furthermore, both
morphologies present a luminosity-dependent bias, with the early types
showing a slightly stronger dependence on the luminosity. We also find
that red galaxies are significantly more clustered than blue ones, with
a mean relative bias of $b_R/b_B$ $\sim$ 1.4, stronger than that seen for
morphology. Finally, by comparing our results with the measurements
obtained from the infrared-selected galaxies we determine that the
relative bias between optical and \iras galaxies in real space is
$b_o/b_I$ $\sim$ 1.4. 

\end{abstract}
 
\clearpage

\section {Introduction}

Optical surveys using full sampling (CfA2, Geller \& Huchra, 1989;
SSRS2, da Costa \etal 1994, 1997) now probe  over 30\% of the sky down
to a magnitude limit of $m_B$ = 15.5. These surveys allow delineating
various large-scale structures  in the different volumes, thanks to
their wide angular coverage. The combination of the large number of
galaxies, complete sampling and sky coverage makes these surveys
extremely useful to study some of the statistical properties of galaxy
clustering. Thus, it becomes possible to divide the sample in bins of
luminosities, morphologies and colors and study the dependence of
clustering on these parameters.

The simplest characterization of the galaxy clustering can be expressed
in terms of the two-point correlation function $\xi$. This statistic
has been widely applied to a variety of samples which include surveys of
optical (\eg Davis \&
Peebles 1983; Davis et al. 1988; de Lapparent et al. 1988; Pellegrini
\etal 1990$b$; Santiago \& da Costa 1990; Loveday 
\etal 1995; Marzke \etal 1995; Hermit \etal 1996; Tucker \etal 1997;
Guzzo et al. 1997) and \iras galaxies (\eg Davis \etal 1988; Saunders
et al. 1992; Strauss et 
al. 1992; Fisher et al. 1994).   The two-point correlation function has
also been used to characterize the dependence of the galaxy clustering
on the internal properties of galaxies such as morphology (\eg Davis \&
Geller 1976; Giovanelli, Haynes \& Chincarini 1986; Iovino et al. 1993;
Loveday et al. 1995; Hermit et al. 1996; Guzzo et al. 1997), color
(Tucker et al. 1996), surface brightness (Santiago \& da Costa 1990),
luminosity (\eg Benoist et al. 1996 and references therein; Valotto \&
Lambas 1997; Guzzo et al. 1997) and internal dynamics (White, Tully \&
Davis 1988). However, since most samples analyzed so far have been 
relatively small, the quantitative results are by and large tentative.

In this paper we use the SSRS2 to investigate the
correlation properties of galaxies in the nearby Universe, comparing our
results with those obtained in other surveys such as the Stromlo-APM
(Loveday \etal 1995), Las Campanas Redshift Survey (LCRS, Tucker \etal
1997) and \iras (Fisher \etal 1994).  We also measure how sensitive
our results are to sample-to-sample variations. With our enlarged
sample, we re-examine the luminosity dependence of $\xi$, previously
studied by Benoist et al. (1996) who only used the southern galactic
cap portion of the SSRS2. Finally, we investigate the dependence of the
correlation properties on morphologies and colors. This information 
is a key ingredient for studies of galaxy biasing.
 
In Section 2 we describe the different catalogs we analyzed, while in
Section 3 we describe the method used to compute the two-point
correlation function in redshift and real space. In Section 4 we
present the results we obtain for magnitude-limited samples. In
Section 5 we examine the dependence of clustering on internal
properties of galaxies, while in Section 6 we compare the correlation
properties of our optical sample with the infrared-selected \iras 1.2 Jy
survey (Fisher et al. 1994). A summary is presented in Section 7. 
 
\section {Data}
 
In this work we use data obtained for the SSRS2, which contains 5512
galaxies in both galactic hemispheres. The data are discussed in greater
detail by da Costa \etal (1994; 1997) so only a brief description will
be made here. The SSRS2 is extracted from the list of non-stellar
objects in the Hubble Space Telescope Guide Star Catalog (Lasker \etal
1990, hereafter GSC).  The SSRS2 is a complete catalog which is
magnitude-limited at $m_B$ = 15.5. Alonso \etal (1994) have shown that
the galaxy magnitudes derived from the GSC are on a uniform isophotal
magnitude system ($\sim$ 26 $mag$ $arc$ $sec^{-2}$) with estimated
magnitude errors \ap 0.3 $mag$. The SSRS2 south contains 3573 galaxies
distributed over 1.13 steradians of the southern galactic cap ($b \leq
-40^o$), within the declination range $-40^o \leq \delta \leq -2.5^o$.
The SSRS2 north contains 1939 galaxies distributed over a solid angle of
0.57 steradians with  $\delta \leq 0^o$ with $b \geq +35^o$. As
described by da Costa \etal (1997) the galaxy morphologies in this
sample used classifications based on other works (\eg Lauberts \&
Valentijn 1989) as well as those made by the authors.

In this work we also considered a sub-sample of SSRS2 galaxies which have
measured colors. For this we used the Lauberts \& Valentijn (1989)
catalog, so that the subcatalog of
galaxies with colors only contains objects south of $\delta$ $\approx$
-17.5$^o$, in both galactic hemispheres. We also imposed a cut at $m_B$
= 14.5 because beyond this magnitude the Lauberts \& Valentijn 
(1989) catalog becomes incomplete, in particular for early type galaxies
(Pellegrini \etal 1990$a$). This selection effect is caused by the fact
that the Lauberts \& Valentijn (1989) catalog is derived from the
diameter-limited Lauberts (1982) catalog. The sample limited at $m_B$=14.5
contains 780 galaxies in both galactic hemispheres, of which 694
(89\%) have colors.  As pointed out by 
Marzke \& da Costa (1997), this sample presents no systematic dependence of the
color completeness with magnitude down to the 14.5 limit of this
sub-sample.

All galaxy heliocentric velocities, $v_\odot$,  have been 
corrected for the Solar motion with respect to the centroid of the Local 
Group using  $v =v_\odot + 300 sin(l) cos(b)$ \kms, where ($l$, $b$)
are the galactic coordinates of the galaxy. We also remove 
from the analyses
all galaxies with radial velocities less that 500 \kms, as the
redshifts for these objects are likely to be dominated by their peculiar
velocities. We use throughout $H_o = 100 h$\kms Mpc$^{-1}$.

\section{The Two-Point Correlation Function}
\subsection{Method}

The two-point correlation function $\xi (r)$ can be computed from the
data using the estimator suggested by Hamilton (1993):  
\begin{equation}
\xi(r) = {DD(r) RR(r) \over [DR(r)]^2}  - 1,
\end{equation}
where $DD(r)$, $RR(r)$ and  $DR(r)$ are the number of data--data,
random--random and 
data--random pairs, with separations in the interval between $r$ and $r +
dr$.  The random catalog is generated using the same selection criteria
as the galaxy sample. This estimator has the advantage that it is not
too sensitive to uncertainties in the mean density, which is only a
second order effect.
The counts $DD(r)$, $DR(r)$ and $RR(r)$ can be generalized to include a
weight $w$ which is 
particularly important to correct for selection effects at large
distances in magnitude-limited samples:
\begin{eqnarray}
DD(r) = &{\displaystyle{ \sum_i^{N_{\scriptscriptstyle{gal}}}
        \sum_j^{N_{\scriptscriptstyle{gal}}}} } & w(s_j, r) w(s_i, r), \\
        & \scriptscriptstyle {r- \Delta r \leq |s_i-s_j| \leq r+
\Delta r} \nonumber
\end{eqnarray}
where $i$ sums over all objects in the sample and the sum over $j$ includes
all particles at a distance $s$ from the origin, which in this work is
taken as the centroid of the Local Group, and  $r  = | {\bf{s}}_i
-{\bf{s}}_j|$ is the separation of the pair $(i,j)$. The
galaxy-random pairs $DR(r)$ and random-random pairs $RR(r)$ are
similarly weighted. The most common weighting schemes are:  equally
weighted pairs $w(s_i, r)=1$; equally--weighted volumes where 
$w(s_i, r)=1/\phi(s_i)$ and the minimum variance weighting given by 
\begin{equation}
w (s_i, r) = {1 \over 1 +4\pi \bar n J_3(r) \phi(s_i)}, \quad J_3 (r)
= \int_0^r dr'r'^2 \xi(r'),
\end{equation}
where $\phi (s_i)$ is the selection
function at distance $s_i$ from the origin and $J_3$ is the
mean number of excess galaxies out to a distance $r$ around each galaxy.
Even though in the last scheme the weights depend on the
unknown correlation function, in practice, it is not very sensitive to
the exact form of $\xi(r)$ (\eg Loveday \etal 1992; Marzke, Huchra \&
Geller 1994). In this work we adopt the minimum-variance
weighting and take  $J_3 (r$ = 30 \h1 Mpc) $\sim 1100$, obtained by
using the real-space correlation function of Davis \& Peebles (1983).
The mean densities were calculated using the estimator 
\begin{equation}
\bar n = \sum_{i=1}^{N_{gal}}w_i/\int_{s_{min}}^{s_{max}} dV\phi(s)w(s)
\end{equation}
where again  $\phi(s)$ is the selection function, derived from the
luminosity function and $w(s)$ is the
weight (\eg Davis \& Huchra 1982). 
The errors for the redshift space correlation function (\xis) as well
as for the real-space correlation 
discussed below, were calculated by means of bootstrap resampling
(Ling, Frenk \& Barrow 1986). For the volume-limited samples the total
of bootstraps was 50, while for magnitude-limited samples 25
resamplings were calculated. As shown by Fisher \etal (1994),
bootstraping tends to overestimate the true errors, so that the
estimate of the latter will in general be rather conservative.

\subsection{Real space} 

In order to estimate real space correlation functions, we follow Davis
\& Peebles (1983). For any two galaxies with
redshifts {\bf s$_1$} and {\bf s$_2$}, we define the separation in
redshift space, and the separation
perpendicular to the line of sight respectively as 
\begin{equation}
{\bf {s = s_1 - s_2}}, \quad {\bf {l}} = {1 \over 2} \bf{(s_1 + s_2)}, 
\end{equation} 
in the small angle approximation.
From these parameters one can derive $\pi$, the separation between two
galaxies parallel to the line of sight and $r_p$, the separation
perpendicular to the line of sight using:
\begin{equation}
\pi = {{\bf s.l} \over |l|}, \quad
r_p = \sqrt{|{\bf{s}}|^2 -\pi^2}.
\end{equation}
These are then used to compute the statistic $\xi(r_p,\pi)$ estimated
from the pair-counts as 
\begin{equation}
1 + \xi(r_p,\pi) = {DD(r_p,\pi) RR(r_p,\pi) \over [DR(r_p,\pi)]^2}.
\end{equation} 
 
From \xip we define the projected function:
\begin{equation}
\omega(r_p) = 2 \int_0^\infty d\pi \quad \xi (r_p,\pi),
\end{equation}
which is related to the real space correlation function through
\begin{equation}
\omega(r_p) = 2\int_0^\infty dy \quad \xi[(r_p^2 + y^2)^{1/2}].
\end{equation}
The inverse is the Abel integral:
\begin{equation}
\xi(r) = -{1 \over \pi} \int_r^\infty dr_p {\omega^\prime(r_p) \over (r_p^2-r^2)^{1/2},
} \end{equation}
where $\omega^\prime(r_p)$ is the first derivative of $\omega
(r_p)$. If the real space correlation function
is a power-law, the integral for $\omega(r_p)$ can be performed
analytically to give 
\begin{equation}
\omega(r_p) = r_p ({r_o \over r_p})^{\gamma} { \Gamma({1 \over 2}) \Gamma({\gamma -1
\over 2}) \over \Gamma({\gamma \over 2})}.
\end{equation}

\subsection {Biasing}
 
The variance of galaxy counts measures the clustering amplitude at
intermediate scales. It is also a useful quantity to compare models 
and data.  The variance in the counts is defined as 
\begin{equation}
\langle (N -nV)^2 \rangle = nV +n^2V^2\sigma^2,
\end{equation}
where nV is the mean number of galaxies in the volume V and $
n^2V^2\sigma^2$ is the mean number of galaxies in excess of random
inside a sphere of volume V.  It is related
to the moment of the correlation function (Peebles 1980)
\begin{equation}
\sigma^2 = {1 \over V^2} \int_VdV_1dV_2 \xi(|r_1-r_2|),
\end{equation}
which can be calculated numerically. For a power law correlation function
$\xi(r) = (r/r_o)^\gamma$,  and a
spherical volume of radius R we get
\begin{equation}
\sigma^2(R) = 72(r_o/R)^\gamma/\lbrack 2^\gamma(3-\gamma)(4-\gamma)(6-\gamma)\rbrack.
\end{equation}
This is the expression we have used to compute \sig8 , and which is
often used to normalize theoretical models.

The relative bias between two different samples at a given separation
$s$ may be estimated
through (\eg Benoist et al. 1996) :
\begin{equation}
\frac {b} {b_*}(s) = \sqrt{\frac{\xi(s)}{\xi_*(s)}} =
\sqrt{\frac{J_3(s)}{J_3*(s)}},
\end{equation}
where the starred symbols denote a sample taken as a fiducial. 
The relative bias of the clustering may also be estimated through 
\begin{equation}
\frac {b} {b_*}(s) = \sqrt{\frac{\sigma^2(s)}{\sigma_*^2(s)}},
\end{equation}
where $\sigma^2$ is the variance of counts in cells described above.

These are the expressions used in this work to
calculate the relative bias between galaxies of different luminosities
relative to $L^*$ galaxies, as well as for different morphological types
and colors.

\section {Magnitude-limited Samples}

In order to estimate the effects due to the finite volume we are
probing and to estimate the importance of cosmic variance, we
compare the clustering properties of the individual SSRS2 south and
north samples as well as the combined sample, with previous estimates
of $\xi$.
In this analysis, we have computed \xis 
taking into account all galaxies brighter than $M=-13$ in 
the velocity range $500 < v < 12,000$ \kms. 
The correlation function was computed using the minimum-variance
weighting discussed in Section 3 and a random background catalog of
10,000 points for the individual
samples and 20,000 points for the combined sample.
In the calculation of the selection function we have used the Schechter
parameters determined for the entire SSRS2 survey by da Costa et
al. (1997), 
which are $M^*$ = -19.55 and $\alpha$ = -1.15. These values are
virtually identical to those measured by da Costa et al. (1994) for
the SSRS2 south.

We tested whether our results are affected by the presence of
clusters of galaxies. For this we used a list of galaxy clusters with
richness  R $\geq$ 1 (J. Huchra, private communication). All galaxies
whose positions were within one Abell radius of the central position
of cluster, and that had radial velocities  within 500 \kms of the
cluster's mean radial velocity were culled from the sample. 
We find that the correlation parameters are virtually unchanged for the
vast majority of the samples, and when there are changes, these are
within the quoted errors of the complete sample. Therefore, we will
not consider the removal of galaxies in clusters in this work.

The redshift space correlation function, \xis, for the SSRS2 samples is
shown in Fig. 1, where we plot the correlation function out to
separations of 30 \h1 Mpc. For the sake of clarity, in the figure we
only show error bars calculated for the combined sample.
 One can see that 
beyond $\sim$ 15 \h1 Mpc,
the errors become progressively larger, and sometimes the
sample-to-sample variations are larger than the estimated errors. In
general, \xis is adequately described by a power-law on small scales.
For most cases in this paper, the power-law fits were calculated in the
interval $2 < s < 11$ \h1 Mpc. The upper-limit was chosen because there
is a suggestion of an abrupt change of slope in \xis on scales $s$
$\lsim$ 12 \h1 Mpc. The lower-limit was chosen to minimize the effects
on \xis due to  peculiar motions of galaxies in virialized systems. The
best power-law fits obtained for each sub-sample of the SSRS2 are
represented as lines in Fig. 1, as explained in the caption. The
correlation parameters derived from the fits are presented in Table 1,
where we list:  the sample identification (column 1); the correlation
length (column 2) and slope $\gamma_s$ (column 3) obtained from the
power-law fits; and in column (4) the rms variance in galaxy counts
within spheres  8 \h1 Mpc in radius, followed in columns (5) through (7)
by the same parameters determined for real space, which will be
discussed below.

An inspection of both Table 1 and Fig. 1 shows that the redshift
correlation functions for SSRS2 sub-samples are very similar on small
scales ($s < 10$ \h1 Mpc). This also demonstrates that the sampling
variations are consistent with the error estimates, at least in the
range of separations for which the fits are calculated. In Fig. 2 we
compare \xis measured in this work for the combined SSRS2 with the \xis
measured in other surveys - the sparsely sampled Stromlo-APM survey
(Loveday et al. 1995), the Las Campanas Redshift Survey (LCRS) (Tucker
\etal 1997), and that measured by Fisher et al. (1994) for the 1.2 Jy
\iras survey. The fit parameters calculated in these papers, as well as
by other workers can be found in Table 2.  Despite small differences in
amplitude, the shapes of the three optical surveys are remarkably
similar. It is important to note that the volumes of the Stromlo-APM (
2.5 $\times$ 10$^6$ $h^{-3}$ Mpc$^3$) and the LCRS ( 2.6 $\times$ 10$^6$
$h^{-3}$ Mpc$^3$) are about 5 times larger than that of the SSRS2 (5.2
$\times$ 10$^5$ $h^{-3}$ Mpc$^3$), and probe different regions of space,
and thus independent structures. The lower amplitude of the \iras survey
compared to the optical samples reflects the relative bias that exists
between optically and infrared-selected galaxies, which will be further
discussed in Section 6 below.

In Fig. 3, we compare our power-law fit parameters with equivalent
measurements by other authors (see Table 2). In general, there is a
good agreement between our values for the redshift
space parameters and those obtained from other optical surveys,
specially the Stromlo-APM and LCRS.

The effect of redshift distortions on the observed redshift correlation
function has also been estimated for the SSRS2. These distortions are
caused by the peculiar velocities of galaxies, which on large scales,
are due to the infall of galaxies from low-density regions into
high-density regions, while on small scales the correlations are smeared
out by virial motions of galaxies in groups and clusters (\eg Kaiser
1987). As described in Section 3, these effects may be accounted for by
calculating the correlation function as a function of the separations
parallel and perpendicular to the line of sight, which can then be used
to define the \wrp estimator, which is unaffected by redshift
distortions. However, one should bear in mind that in general the
calculation of the real space correlation function is much more
susceptible to noise than that calculated in redshift space.

From the correlation functions \xip computed using the minimum-variance
weighting scheme, we have obtained \wrp. From power-law fits, in the
interval 2 $< r_p <$10 \h1 Mpc,  we have derived the correlation
parameters listed in Table 1. By comparing the real space fit parameters
 obtained in this work (Table 1) with previous measures (columns 4 and 5
in Table 2), we find a good agreement with the real space measurements
of Davis \& Peebles (1983), Loveday et al. (1995) and Marzke et al.
(1995). 

The fit to the real space correlation function for the combined
sample is compared in Fig. 4 with the redshift space correlation function.
One can see that at intermediate separations the redshift \xis is
amplified relative to the real space correlation \xir. The small
amplification suggests that the observed redshift distribution is close
to the real space distribution. At separations of $\sim$ 10 \h1 Mpc, the
ratio between the real and redshift space correlations is $\sim$ 1.5. In
the linear regime, peculiar motions on large scales cause \xir to be
amplified by a factor $\sim$ $1 + { 1 \over 2 } {\beta} + { 1 \over 5 }
{\beta^2}$ where $\beta = {\Omega^{0.6} \over b}$ and $b$ is the linear
biasing factor (Kaiser 1987). Therefore, a rough estimate for $\beta$ is
$ \sim 0.6$, on scales of the order of 10 \h1 Mpc, consistent
with that determined by Loveday \etal (1996).

\section {The Clustering Dependence on the Internal Properties of Galaxies}

\subsection {Luminosity}

In this work we use the combined SSRS2, as well as the SSRS2 north and
south sub-samples to further explore the clustering dependence on
luminosity, as was carried out by Benoist et al. (1996), but who only
used the SSRS2 south. 
Probing independent structures in different volumes we can estimate 
the impact of cosmic variance. It should also be noted that the
absolute magnitude limits considered in this section differ slightly
from those of Benoist et al. (1996), and were chosen to
compare our results with the volume-limited samples of Fisher \etal
(1994), which will be discussed in Section 6 below.

The volume-limited samples considered in this section only contain
galaxies bright enough that would allow them to be included in the
sample when placed at the cutoff distance. We defined samples limited at
radial distances of 60, 80, 100 and 120 \h1 Mpc. The absolute magnitude
limits corresponding to these distances are  -18.39, -19.01 (both $L <
L^*$) , -19.50 ($\sim L^*$)  and -19.89 ($L > L^*$), respectively. For
all galaxies in these samples, the weighting function is $w(r)=1$ and
the volume densities  are simply the total number of galaxies divided by
the corresponding volume. 

The correlation functions obtained for the volume-limited sub-samples
at the different depths are shown in Fig. 5 for $s \leq 20$ \h1 Mpc,
where the different symbols represent different volume limits. For
reasons of clarity, we only present error bars for the samples
volume-limited at 60 \h1 and 120 \h1 Mpc. The meaning of these
symbols, as well as the indication of the parent sample (SSRS2 south,
north or combined) are shown in each panel. The power-law fits are
represented by lines in the figure and the parameters are summarized
in Table 3 where we list: in column (1) the sample; in column (2) the
depth R; in column (3) the number of galaxies N$_g$; in column (4) the
mean density; in columns (5) and (6) the power-law fit parameters and
formal errors; and in column (7) $\sigma_8$, the rms fluctuation of
the number of galaxies in a sphere of radius 8\h1 Mpc. The interval
used in the fits is $ 2 < s < 11$ \h1 Mpc, the same as that adopted in
the previous section.

An inspection of Table 3 and Fig. 5 shows that the amplitude of \xis
tends to increase with the sample  depth, the variation being somewhat
larger in the northern and combined samples.  We point out that \xis for
the SSRS2 north (panel b), is noisier because of the smaller number of
galaxies, in most cases about half of those in the southern sample.  The
correlation length ($s_0$) ranges from 3.8 \h1 Mpc to 6.8 \h1 Mpc.
However, the slope varies considerably from sample to sample, though
with a tendency of becoming steeper as the depth increases.

In order to evaluate the cosmic variance, we show in Fig. 6 \xis for
each of the volume-limited sub-samples, but  now plotting the results
for the southern, northern and combined samples in each panel. For the
samples in smaller volumes, the differences between the northern and
southern samples are larger than the estimated error calculated for
the combined sample, and probably
reflect the amplitude of the sample to sample variation, with the north
being systematically lower. For the larger volumes the samples present
similar behavior, and the variations are generally consistent with the
estimated errors.

To remove the effects of distortions due to motions,
which may affect our estimates of the strength of clustering and the
relative bias between different samples, we have also computed the
real-space correlation function for the volume-limited samples.
As above, we have computed \xip for the sub-samples volume-limited at 
R = 60, 80, 100  and 120 \h1 Mpc in each
galactic hemisphere and for the combined sample. The resulting real
space correlation parameters are listed in Table 4.

In Fig. 7 we compare \xis  measured for each volume limit, denoted by
open symbols, with the real-space correlation fits described above,
represented as a solid line. For the sake of clarity, we only show the
fits we measure for the combined sample, as this will be the one less
affected by noise. The smearing due to motions in virialized systems for
$r < 3$ \h1 Mpc is quite noticeable for all samples, while the effect of
peculiar motions is only obvious for the smaller volumes, little
evidence being seen in the samples in larger volumes. 

The dependence of clustering in redshift--space (as measured by 
$\sigma_8$) with luminosity (as measured by the limiting absolute 
magnitude of each sample) is shown in Fig. 8(a), where we use as 
fiducial magnitude the value of $M^*$=-19.55 (see Section 4). The
figure shows
an overall behavior consistent with that found by Benoist et
al. (1996) and which is detected in all samples, further
demonstrating that this effect is unlikely to be spurious.
This result supports their finding
that there is a dependence of clustering on luminosity, as measured in
redshift space. To further investigate its reality, we have computed
\xir in real space for the same volume limited samples. The results are
shown in Fig. 8(b). Here again it is immediately apparent that the
clustering amplitude increases with luminosity in the same way as seen
in redshift space. On the whole, these results, using a larger sample,
confirm in real space the conclusions of Benoist \etal (1996).

In Fig. 9 we present the relative bias with scale calculated using
equation (15), where we compare the the correlation function for the
volume-limited samples at 60, 80 and 120 \h1 Mpc relative to the 100 \h1
Mpc sample.
From the figure one may see that there are only minor
differences between the smaller volumes. In the case of the sample
volume-limited at 120 \h1 Mpc, the relative bias is fairly constant over
the range of scales we consider at $\sim$ 1.5. This suggests that the
luminosity bias is scale-independent, and that it starts to become
important only for galaxies brighter than  $\sim$ $L^*$.

\subsection{Morphology}

Since all galaxies in the SSRS2 have morphological classifications we
can also analyze the clustering dependence on morphology. With this aim,
we have calculated \xis and $\xi(r)$ for the SSRS2 for different
morphological types, dividing galaxies into broad morphological  bins -
early types comprising  E, S0 and S0-a, and late types containing Sa
galaxies and later. In contrast to the results of the Stromlo-APM, the
luminosity function parameters used in the selection function for both
samples are quite similar to those measured for the SSRS2 as a whole
(Marzke et al. 1997). Furthermore, since sample-to-sample variations
are within our estimated errors both for the magnitude and
volume-limited samples, as shown in Section 4, in the analysis below
we only consider the combined sample to improve the statistics.

The resulting correlation functions for early and late type galaxies are
presented in Figure 10 panel (a) in redshift space and (b) 
in  real space, while the fit parameters are presented in Table 5. For
the late type galaxies we find that the  correlation function is
adequately described by ($s_0$ = 5.4\mm0.2\h1 Mpc;
$\gamma_s$=1.48\mm0.09), while for early types we find  $s_0$ =
6.5\mm0.2 \h1 Mpc; $\gamma_s$=1.86\mm0.11. Our values for early type
galaxies are close to those of Santiago \& da Costa for the
diameter-limited SSRS ($s_0$ = 6.0\mm1.5 \h1 Mpc, $\gamma_s$=1.69)  and
Hermit \etal (1996) for the ORS, who measure $s_0$=6.7 and
$\gamma_s$=1.52. Our value for late types is somewhat larger than that
measured by Santiago \& da Costa (1990), while a proper comparison  with
Hermit \etal (1996) cannot be made, because we have not subdivided
spirals into earlier (Sa/Sb) and later (Sc/Sd) types as they did. A
comparison between the fit parameters obtained from available redshift
space correlation functions, is shown in Fig. 11, where open symbols
represent fits for late type galaxies and solid symbols represent early
types.  Although all works agree that early types are more clustered than
late types, as indicated by the larger correlation length, the scatter
is large with the Stromlo-APM results yielding very extreme results.
This, in turn, implies large uncertainties in the measurement of the
relative bias between the two populations.  Based on the redshift space
information, we estimate the relative bias between morphological types
as 1.25.

However, for a proper estimate of the dependence of the correlation
properties on morphology it is important to take into account the fact
that redshift distortions may affect early and late type galaxies in
different ways. Therefore a more meaningful comparison must be carried
out in real space. The values we measure for the correlation length in
real space for early types ($r_0$=6.0\mm0.4; $\gamma_r$=1.91\mm0.18)
show a very good  agreement with those of Loveday \etal (1995),
($r_0$=5.9\mm0.7; $\gamma_r$=1.85\mm0.13). For late types we find
($r_0$=5.3\mm0.3; $\gamma_r$=1.89\mm0.15), which is somewhat larger 
than those measured by the same authors ($r_0$=4.4\mm0.1;
$\gamma_r$=1.64\mm0.05).

Our value of the correlation length is significantly smaller than that
measured by Guzzo et al. (1997) ($r_0$=8.4\mm0.8; $\gamma_r$=2.05\mm0.09)
for early types. We should note that their sample is volume-limited at
$M < -19.5$, whereas we consider galaxies down to $M = -13$.  In order
to compare with these authors we consider a volume-limited sub-sample of
SSRS2 galaxies with $M \leq$ -19.5, which corresponds to maximum
distance of 100 \h1 Mpc. Using this sample, for early types we measure
($r_0$=5.7\mm0.8; $\gamma_r$=2.09\mm0.49) while for late types we find
($r_0$=5.0\mm0.5; $\gamma_r$=2.01\mm0.28).  For both early as well as
late types, there are still discrepancies relative to the results of
Guzzo et al. (1997), which could reflect the paucity of rich clusters in
our sample.

By using the variance, we estimate that the relative bias between the
different morphologies is $b_{E+S0}/b_S$ = 1.18$\pm$0.15 in a sample
where clusters are not important.  This value is smaller than the
determination derived 
from the real-space correlations of Loveday et al. (1995)  $b_{E+S0}/b_S$ =
1.33 and Guzzo et al. (1997) $b_{E+S0}/b_S$ = 1.68. From these results
we may conclude that the relative bias between the two populations
range from roughly 1.2 to 1.7, depending on the cluster abundance
in the sample, with the former value representing a lower limit.

We have also calculated the correlation function for galaxies
discriminated by morphological types for volume-limited samples using
the same absolute magnitude limits as in Section 5.1. This calculation
was carried out both in redshift, as well as real space, and the results
are presented in Tables 6 and 7 respectively.  In redshift space there
is a trend of the correlation function amplitude  increasing with
luminosity for both morphological classes.  The magnitude of this
variation is larger for early types than for late types, although the
errors are large. The same trend may be inferred from the analysis in
real space, as shown in Figure 12(a), where we compare the $\sigma_8$
values obtained for the different sub-samples. Here it may be clearly
seen that there is a trend of $\sigma_8$ increasing with luminosity, 
suggesting that the morphological and luminosity segregations are two
separate effects.

Using equation (15) we can also examine how the relative bias varies as
a function of scale. This is shown in Figure 12(b), using the real space
correlation functions. In contrast to the luminosity bias we find that
the morphological bias presents a small decrease from $\sim$ 1.4 on small
scales to $\sim$ 1.0 on larger scales ($\sim$ 8 \h1 Mpc). Although
the latter value is slightly smaller than that estimated through the
$\sigma_8$ values ($b_{E+S0}/b_S$ = 1.18 $\pm$ 0.15), it is still
within the estimated error. A similar behavior of the morphological
bias changing with scale, was found by Hermit et al. (1996) but using
the redshift space correlation function of the ORS, which may not be
as meaningful, because of possible biases introduced by virial
motions.

Taken together, the above results are consistent with the interpretation
that luminosity segregation could be a primordial effect, while the
morphological segregation could be enhanced by environmental effects
(e.g. Loveday \etal 1995).

\subsection{Colors}

Another internal characteristic available in the present catalog
is color. Although morphology and colors are correlated the scatter is 
large, and galaxies of a given type exhibit a broad range of colors,
indicating different star-formation histories.
On the other hand, colors are easily measured and are
an objective criterion, in particular for
samples of distant galaxies, whereas the morphological classification
is somewhat subjective and becomes increasingly difficult to carry out
as the apparent sizes of galaxies get smaller. A further evidence that
morphology and colors have somewhat different distributions comes from
the calculation of the luminosity function, which presents
significantly different shapes for blue and red galaxies (Marzke \& da
Costa 1997), while the luminosity function calculated by separating
galaxies between early and late types presents similar Schechter
parameters (Marzke et al. 1997).

The few works calculating the correlation properties of galaxies
divided by colors present rather conflicting results for the deep
samples. Works by Infante \& Pritchet (1993) and Landy, Szalay \& Koo
(1996) using the angular correlation function show that the 
correlation of
redder galaxies is significantly stronger than for bluer galaxies,
except for the very bluest ones (Landy et al. 1996). Carlberg et
al. (1996) analyzing a redshift survey of K-band selected galaxies,
find that for $0.3 \leq z \leq 0.9$ red galaxies are more correlated
than blue galaxies by a factor of five. These results differ from those
of Le F\`evre et al. (1996) who find that at $z \geq$ 0.5 blue and red
galaxies have the same correlation properties, while for 0.2 $\leq z
\leq$ 0.5 blue galaxies are less correlated than red ones.
For nearby
galaxies, Tucker et al. (1996) have calculated the correlation function
and showed that at small scales ($s \leq 10$ \h1 Mpc) red galaxies
($[b_J - R]_0 > 1.25$) cluster more strongly than blue ($[b_J - R]_0 <
1.05$) ones, while for larger scales no evidence of color segregation
is seen.

In order to make an independent estimation of the dependence of \xis on
colors, we use the the $m_B$ = 14.5 sample described in Section 2, which
contains galaxies in both galactic hemispheres. As mentioned in Section
2, this bright limit was used because of incompleteness in colors, as we
are restricted to galaxies with measurements in the Lauberts \&
Valentijn (1989) catalog.  In this work we adopted the restframe color
cutoff as $(B_T-R_T)_0$ = 1.3  which is roughly the color of an Sbc
galaxy, and was the criterion adopted 
by Marzke \& da Costa (1997) in
the determination of the luminosity function by colors. This value is
close to the median value of $B_T-R_T$ in our sample which is
$B_T-R_T$=1.2. The conversion of observed into restframe colors used the
no-evolution models calculated by Bruzual \& Charlot (1993), where we
assume that the B and R measures in the Lauberts \& Valentijn (1989)
catalog are on the same system of $b_J$ and $r_F$ used by Bruzual \&
Charlot (1993). To calculate \xis we used the following Schechter
function parameters; for blue galaxies ($B_T-R_T \leq 1.3$), $M^*$ =
-19.43, $\alpha$ = -1.46; for red galaxies ($B_T-R_T > 1.3$), $M^*$ =
-19.25, $\alpha$ = -0.73, which were obtained by Marzke \& da Costa
(1997). The sample, which only considers galaxies out to a maximum
distance of 8000 \kms, contains 387 blue and 219 red galaxies. 

The results of the two-point correlation function are shown in Figure 13
(a) for redshift space while the fit  parameters may be found in Table
8.  Because of the small number of objects, the correlation function is
very noisy, yet it is unquestionable that the red galaxies present a
systematically higher amplitude at all separations compared to blue
galaxies. In order to verify how sensitive the results may be to
incompleteness, we re-calculated \xis for the $m_B$=14.2 sample which is
92 \% complete in colors. The fit parameters present a similar behavior,
although the values differ from those measured for the 14.5 sample. The
results we obtain for the samples discriminated in colors present a
qualitative agreement with those of Tucker \etal  (1996), in the sense
that red galaxies are more strongly correlated than blue galaxies.

We have also calculated the real-space correlation function for the 14.5
sample and the power-law fit is presented in
Fig. 13 (b), together with the redshift space correlation.
The figure shows
that the slopes of both power law fits are fairly  similar
($\gamma_r$=1.99 for blue, $\gamma_r$=2.18 for red  galaxies), though
the uncertainties are rather large, in particular for the red galaxies.
The observed \xir suggests that red galaxies are probably more affected
by peculiar motions than blue galaxies. Because of the relatively small
size of the sample with colors, we have not been able to  investigate
the dependence on luminosity, which would be dominated by errors because
of the small number of objects assigned to each luminosity bin.

The relative bias estimated from $\sigma_8$  in real space is
$b_R$/$b_B$ = 1.40\mm 0.33, and a similar result is obtained if the
redshift space results are considered.  As in the case of luminosity and
morphology, one may calculate the relative bias between galaxies of
different colors as a function of scale, which is presented in Fig. 14.
Because the observed correlation function is rather noisy, for this plot
we used the fits to \xir . Taking the results at face value they would
suggest that the relative bias between red and blue galaxies on small
scales is comparable to that seen for early and late type galaxies.
However, it levels off more rapidly ($\sim $ 4 \h1 Mpc), remaining
constant at $b_R/b_B \sim 1.2$ thereafter. 
This behavior could be the result of evolution due to environmental effects,
where early type galaxies in higher density regions lost their gas more 
rapidly than bluer galaxies, and thus present a much lower star formation
rate. However,  because the errors are large, these results should
only be considered as tentative.

\section{Comparison with \iras galaxies}

In this section we compare the correlation properties of both the entire
SSRS2 sample as well as for the volume-limited sub-samples described in
Section 5.1 with the \iras 1.2 Jy survey (Fisher et  al. 1994). In Fig.
2 we presented a comparison between \xis as  measured by different
surveys. In that figure it is quite apparent that the three optical
surveys are in very good agreement, while the \iras survey presents
systematically lower amplitude, which reflects the bias that exists
between the distribution of  optically-selected and infrared-selected
galaxies, previously noted by several authors (Davis et al 1988; Babul
\& Postman 1990; Lahav et al. 1990; Saunders et al. 1992; Strauss et al.
1992; Fisher et al.  1994). 

The clustering dependence on infrared luminosity was investigated by
Fisher et al. (1994), using different sub-samples of the \iras 1.2 Jy
survey, who found no evidence for such dependence. This result
differs from that presented in Section 5.1 above, and a comparison between
the different volume-limited sub-samples of SSRS2 and \iras galaxies
is  presented in Fig. 15. We point 
out that as there are no available measures in real space calculated by
Fisher \etal (1994), here we use the values obtained in redshift space,
which are presented  in the last column of Table 3. The inspection of
this figure shows that the amplitude and shape of \xis  have the best
agreement for the 60 \h1 Mpc sample ($M < -18.39$), while the other
SSRS2 sub-samples containing brighter galaxies present $\xi(s)_{SSRS2} >
\xi(s)_{IRAS}$. A possible explanation for the different behavior of the
optical and \iras relative to luminosity is that the optical luminosity
is more strongly related to the mass, while the infrared luminosity of
\iras galaxies reflects rather the star-formation rate which is only
weakly dependent on the mass (Davis et al. 1988).

The relative bias between the different volume-limited samples of the
SSRS2 and the \iras 1.2 Jy survey are shown in Fig. 16(a). For this
calculation we used the variance calculated in redshift space, shown in
column (7) of Table 3, with the values in Table 1 of Fisher et al.
(1994). An inspection of the figure shows that the relative bias between
both samples increases with luminosity, ranging from $b_o/b_I$ = 0.94 to
$b_o/b_I$=1.91.  The smallest value is obtained for the sample which
includes less luminous galaxies ($M < -18.39$), while the largest value
is for the sample with the brightest  galaxies ($M < -19.89$). This
result suggests that the relative bias between optical and \iras
galaxies depends on the luminosity of the objects, the less luminous
optical galaxies showing a clustering amplitude comparable to that found
for \iras galaxies.

The mean relative bias between the optical and \iras samples may be
estimated using the $\sigma_8$ values both in redshift and real space.
For this we use the $\sigma_8$ for the magnitude-limited SSRS2 sample
and the 1.2 Jy \iras sample (Fisher \etal 1994), using for the former,
values in the last column of Table 1. By using the value of $\sigma_8$
for the combined sample we find that the relative  bias between optical
and \iras sample is $\sim$ 1.20 \mm 0.07.

The bias in real space  can be obtained using the $\sigma_8$ value
derived from the real space correlation function. For our combined
sample we measure \sig8 = 0.96 $\pm$ 0.06, while Fisher \etal (1994)
quote for \iras galaxies \sig8 = 0.69 $\pm$ 0.04. This result implies in
a relative bias $b_o/b_i = 1.39\pm 0.17$, $\sim$ 17 \%  larger than that
estimated in redshift space. This value is consistent with the value of
$\sigma_8$ = 1.38$\pm$ 0.12 reported by Fisher \etal (1994). 

One may also calculate the relative bias with scale in real space by
using equation (15) for the combined magnitude-limited optical and
flux-limited \iras samples. The results are presented in Fig. 16(b).
Notwithstanding the large error bars, the results suggest
that relative bias between optical and \iras galaxies decrease
with scale varying from about 1.4 on 1 \h1 Mpc to close to 1
on 10 \h1 Mpc scales (e.g. Strauss et al. 1992).

\section {Summary}

We have investigated the correlation properties of galaxies
in the SSRS2 catalog for which we have considered both volume and
magnitude-limited samples. The main results may be summarized as follows:

\begin{itemize}

\item {} In spite of the small volume probed relative to
the scale of inhomogeneities, we find an excellent agreement between our
correlation function and those of other surveys probing volumes more
than 5 times larger.
This result is in contradiction with the fractal interpretation of the galaxy
distribution in the Universe, which predicts that the
correlation length increases with volume. 

\item {} The relatively small differences between  redshift and real
space correlations on intermediate scales ($s \sim$ 10 \h1 Mpc) suggest
a low value of $\beta = \Omega^{0.6}/b < 1$, indicating that the
redshift distribution of galaxies is close to that in real space. 
 
\item {} We confirm the existence of a luminosity-dependent bias for
super-L* galaxies that is scale-independent, suggesting that 
it is of primordial nature.
 
\item {} In contrast, the relative bias between early and late types
shows a scale dependence, varying from about 1.4 on small scales to 1 at
$\sim$ 8 \h1 Mpc. The mean relative bias is found to be  $b_{E+S0}/b_S
\sim$ 1.2. This small value, when compared to  
previous surveys, probably
reflects the paucity of rich clusters in the surveyed region.
 
\item {} Both early and late types show separately a
luminosity-dependent bias similar to the sample as a whole further
suggesting that the luminosity bias is primordial in nature while the
excess clustering of early types relative to spirals on small scales may
be caused by environmental effects.
 
\item {} The relative bias between red and blue galaxies is similar to
that observed between early and late type galaxies. However, it levels
off on smaller scales $\sim$ 4 \h1 Mpc at a constant value of about 1.2.
We find that the mean relative bias of galaxies selected by colors is
greater than when selected by morphologies. We point out, however, that
color samples are significantly smaller and the uncertainties
correspondingly larger.
 
\item {} The relative bias between optical and IRAS galaxies also varies
with scale at least out to $\sim$ 10 \h1 Mpc and shows a strong
luminosity dependence. The mean relative bias between optical and \iras
is $b_o/b_I$ = 1.39 \mm 0.17 in real space. 

\end{itemize}

The results presented here offer key elements for
constraining galaxy formation models. Although intriguing, we should
note that the samples are still relatively small, especially those
with color information, so these results should be considered only as
tentative. Future larger samples are essential to further
investigate these effects, and which are likely to give more insight
on the relation between galaxies and large-scale structures, and on
the galaxy formation process.

\bigskip\bigskip

\noindent  We would like to thank our SSRS2 collaborators for allowing
us to use the data in advance of its publication. We also thank K.
Fisher and J. Loveday for helpful discussions and for providing us with
their results. We thank D. Tucker for providing results of the LCRS
survey and S. Hermit for ORS results. We also thank R. Marzke and M.
Vogeley for many useful discussions. CNAW acknowledges partial support
from CNPq grants 301364/86-9, 453488/96-0 and from the ESO Visitor
program. PSP acknowledges funding from CNPq grant 301373/86-8 and from
the Centro Latino-Americano de F\'\i sica. 
\clearpage

\begin{figure}
\vspace{185mm}
\includegraphics{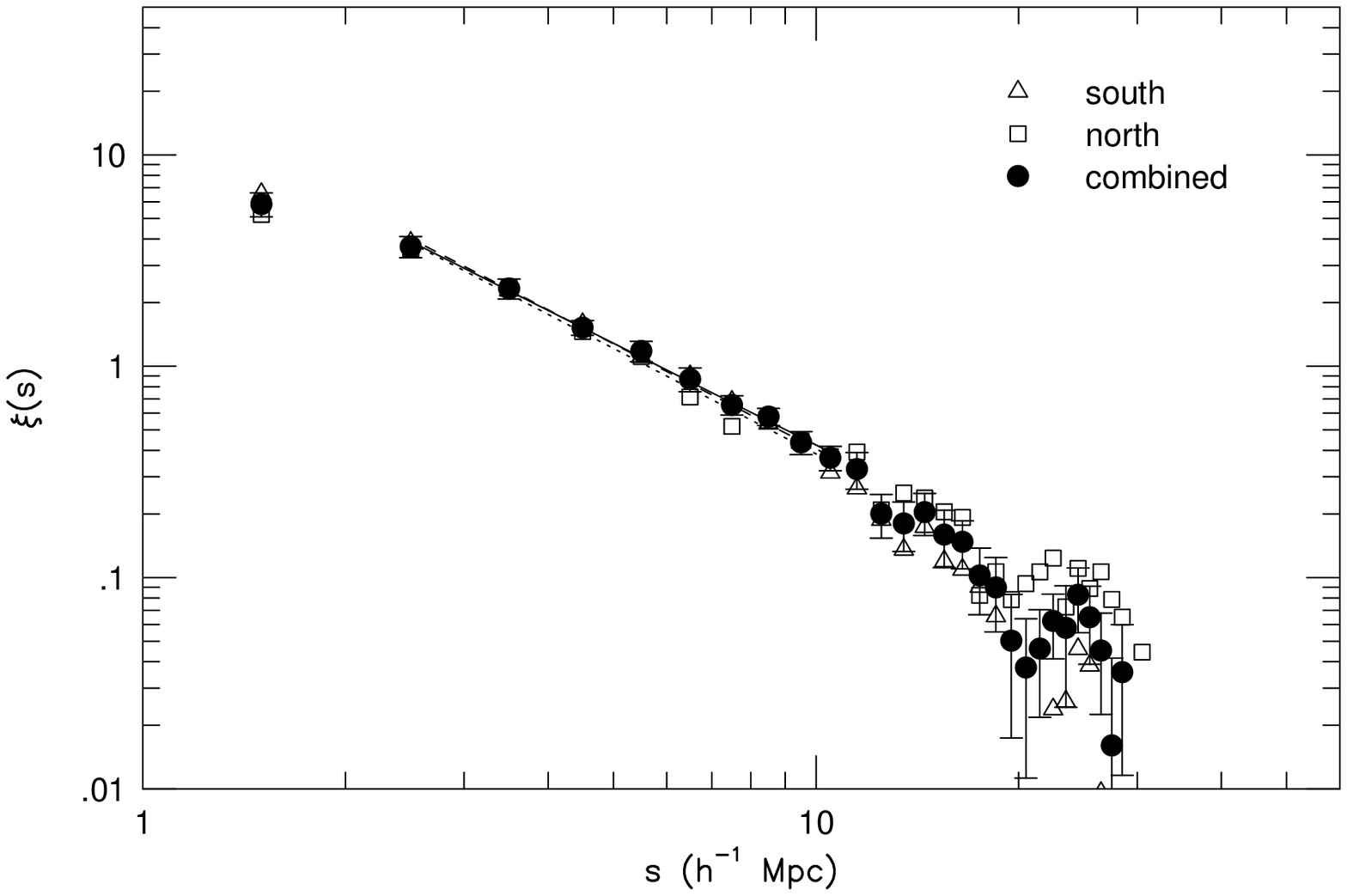}
\caption{ Redshift space correlation function using optimal weighting
and fits for the southern (open triangles; dashes), northern (open
squares; dots) and combined (solid circles; solid line) sub-samples of
the SSRS2. The fit parameters are presented in Table 1. Error bars,
calculated by means of bootstrap resampling are presented only for the
combined sample}
\end {figure}
\clearpage

\begin{figure}
\vspace{185mm}
\includegraphics{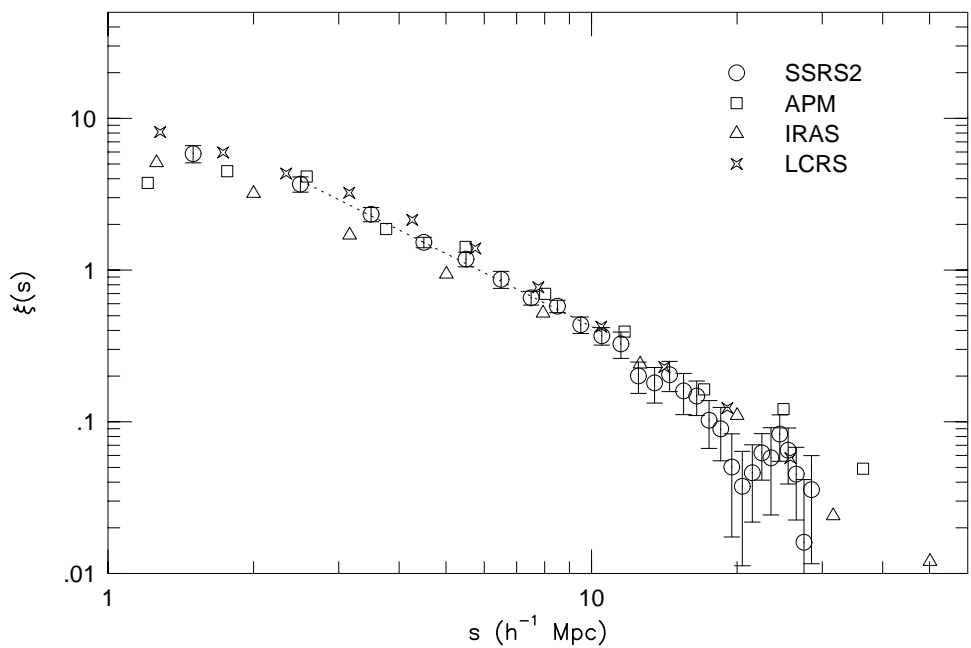}
\caption{Comparison between the redshift correlations of the SSRS2
combined sample (open circles), the optically selected APM (open
squares) and LCRS (crosses) and the infrared-selected $IRAS$ 1.2Jy
(open triangles). The dashed line represents the fit to the SSRS2
correlation function. For the sake of clarity, we only show error bars
calculated for the SSRS2 sample.} 
\end{figure}
\clearpage

\begin{figure}
\vspace{185mm}
\includegraphics{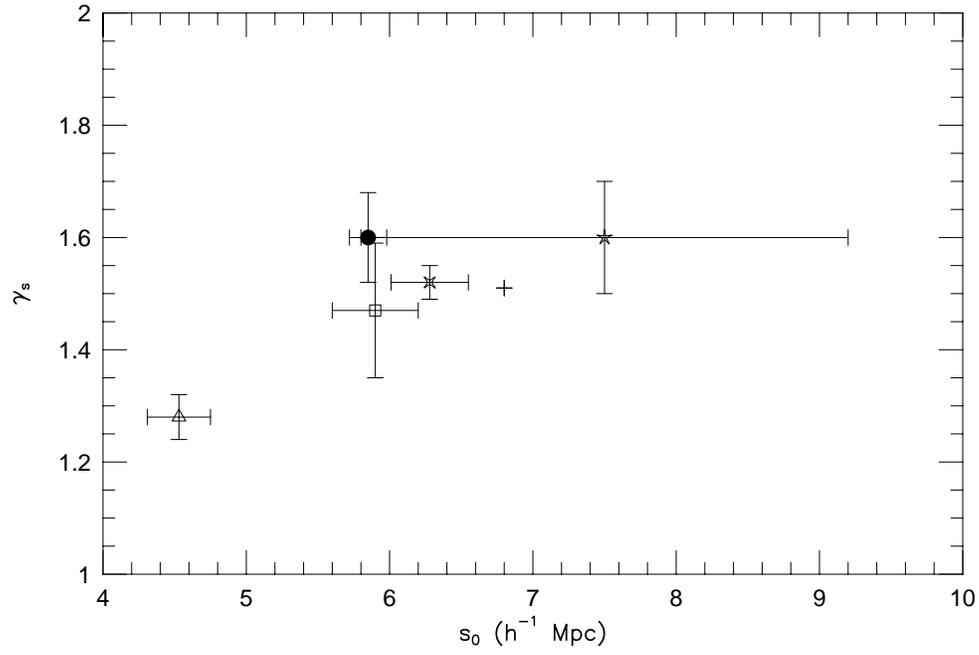}
\caption{Comparison between the correlation function parameters for
the SSRS2 combined sample (solid circle), Stromlo-APM (open
square), LCRS (cross), $IRAS$ 1.2Jy (open triangle), CfA2 (star)
 and ORS (plus). In the case of this last survey, no error
bars were provided.}
\end{figure}
\clearpage

\begin{figure}
\vspace{185mm}
\includegraphics{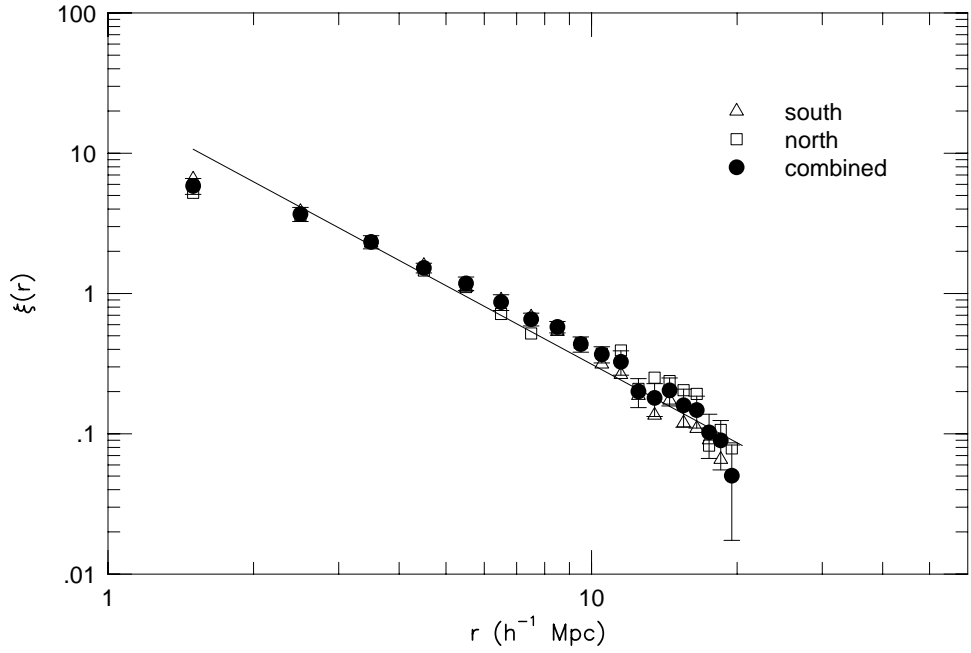}
\caption{Comparison between the real and redshift space correlation functions
calculated for the magnitude-limited north and south sub-samples and combined
SSRS2. The symbols represent redshift space values (as in Fig. 1),
 while the solid line represents the real-space fit for combined
sample. As in Fig. 1, error bars are only shown for the combined sample.
}
\end{figure}
\clearpage

\begin{figure}
\vspace{185mm}
\includegraphics{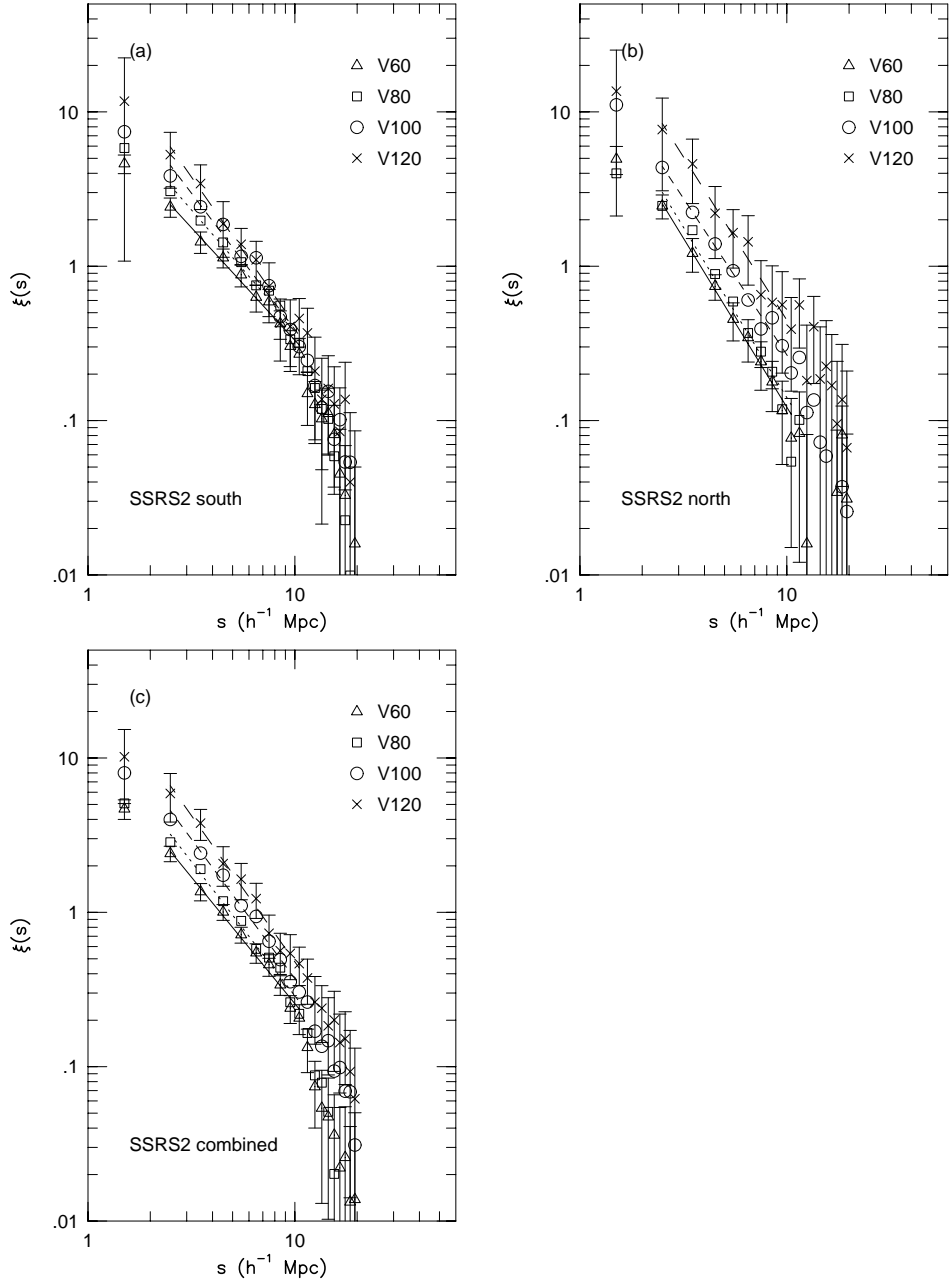}
\caption{ Two-point correlation function in redshift space for the
(a) southern, (b) northern and (c) combined sub-samples of the SSRS2.
The symbols and lines represent the correlation function and fits for
the volume-limited sample: at 60 \h1 Mpc (open
triangles, solid line); 80 \h1 Mpc (open square, dotted line); 100
\h1 Mpc (open circles, short-dashes) and 120 \h1 Mpc (crosses, long
dashes). The fit parameters can be found in Table 3. The bootstrap
error estimates are only shown for the 60 \h1 Mpc and 120 \h1 Mpc
samples for the sake of clarity.}
\end{figure}
\clearpage

\begin{figure}
\vspace{185mm}
\includegraphics{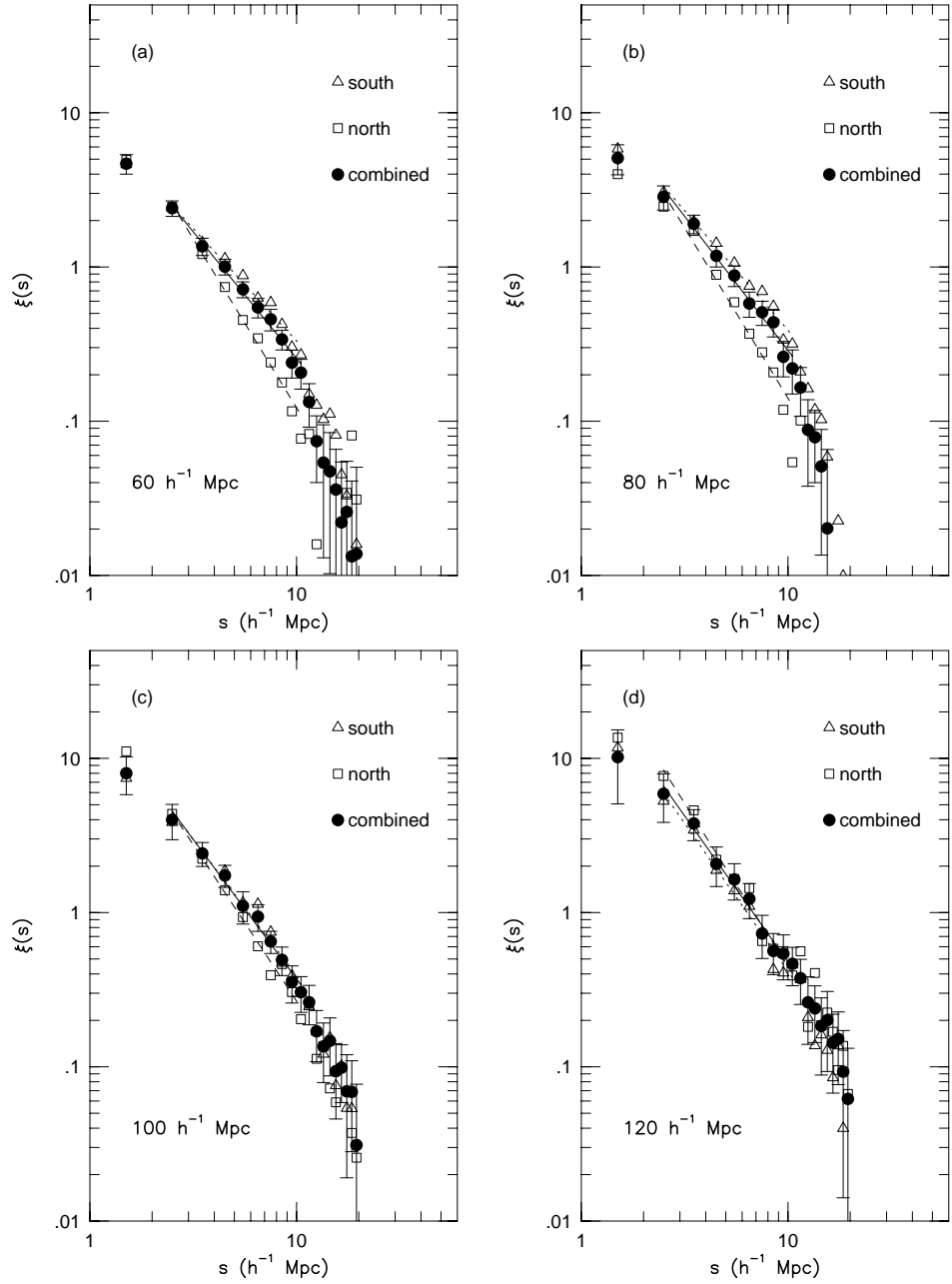}
\caption{ Comparison between the redshift correlation for the
individual and combined samples for each volume limit. The solid line
represents the power-law fit for the combined sample, the dotted line
the fit for the south and the dashed line the fit for the
north. Bootstrap error estimates are shown only for the combined sample.}
\end{figure}
\clearpage

\begin{figure}
\vspace{185mm}
\includegraphics{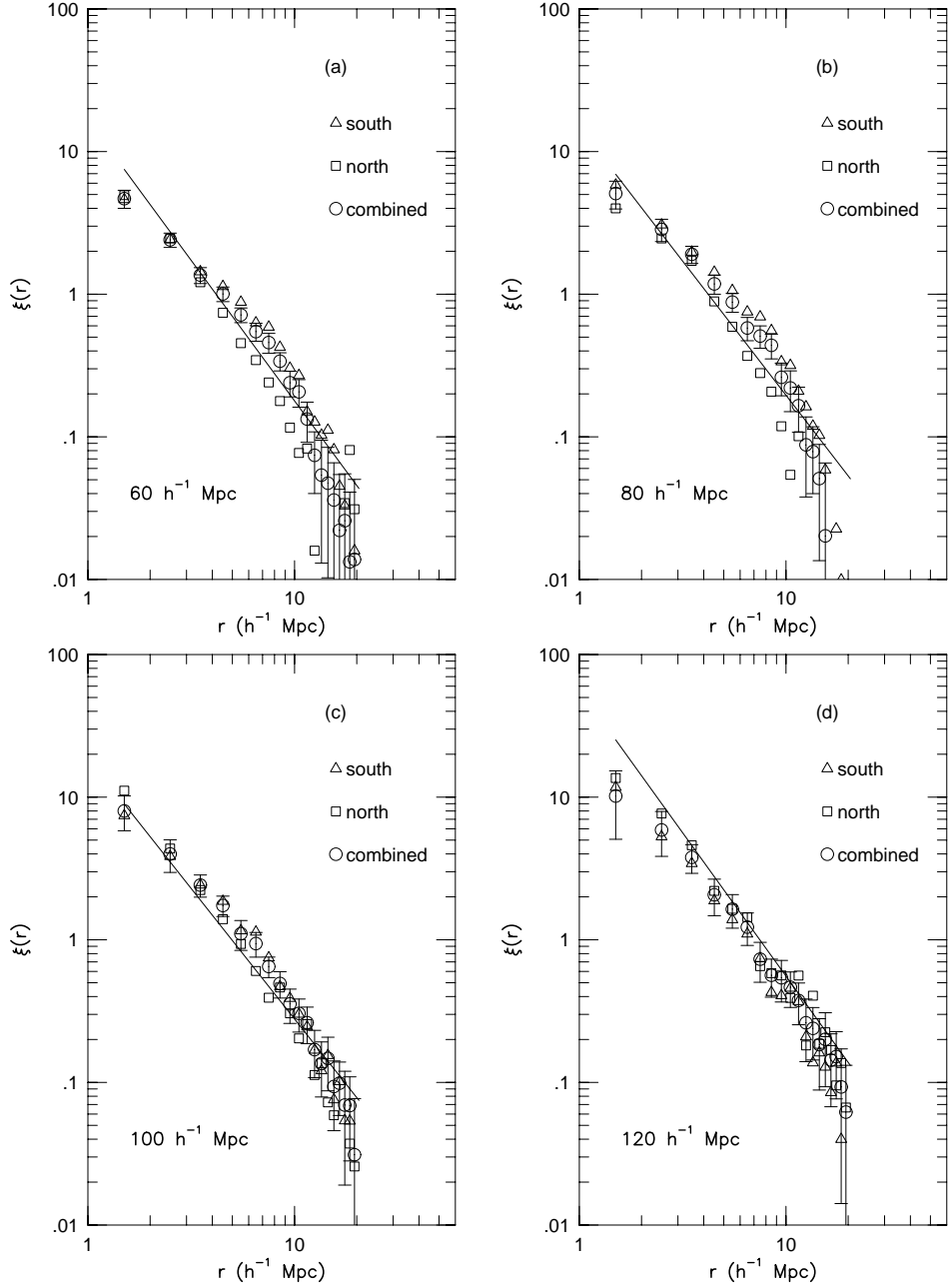}
\caption{Comparison between the redshift space and real space
correlation function, calculated for volume-limited samples.
Open symbols represent the redshift space correlations, while solid
lines represent the fits to the observed $\omega_p(r_p)$ using the
power-law approximation of equation (11) for the combined
sample. Error bars are shown only for the combined sample.} 
\end{figure}
\clearpage

\begin{figure}
\vspace{185mm}
\includegraphics{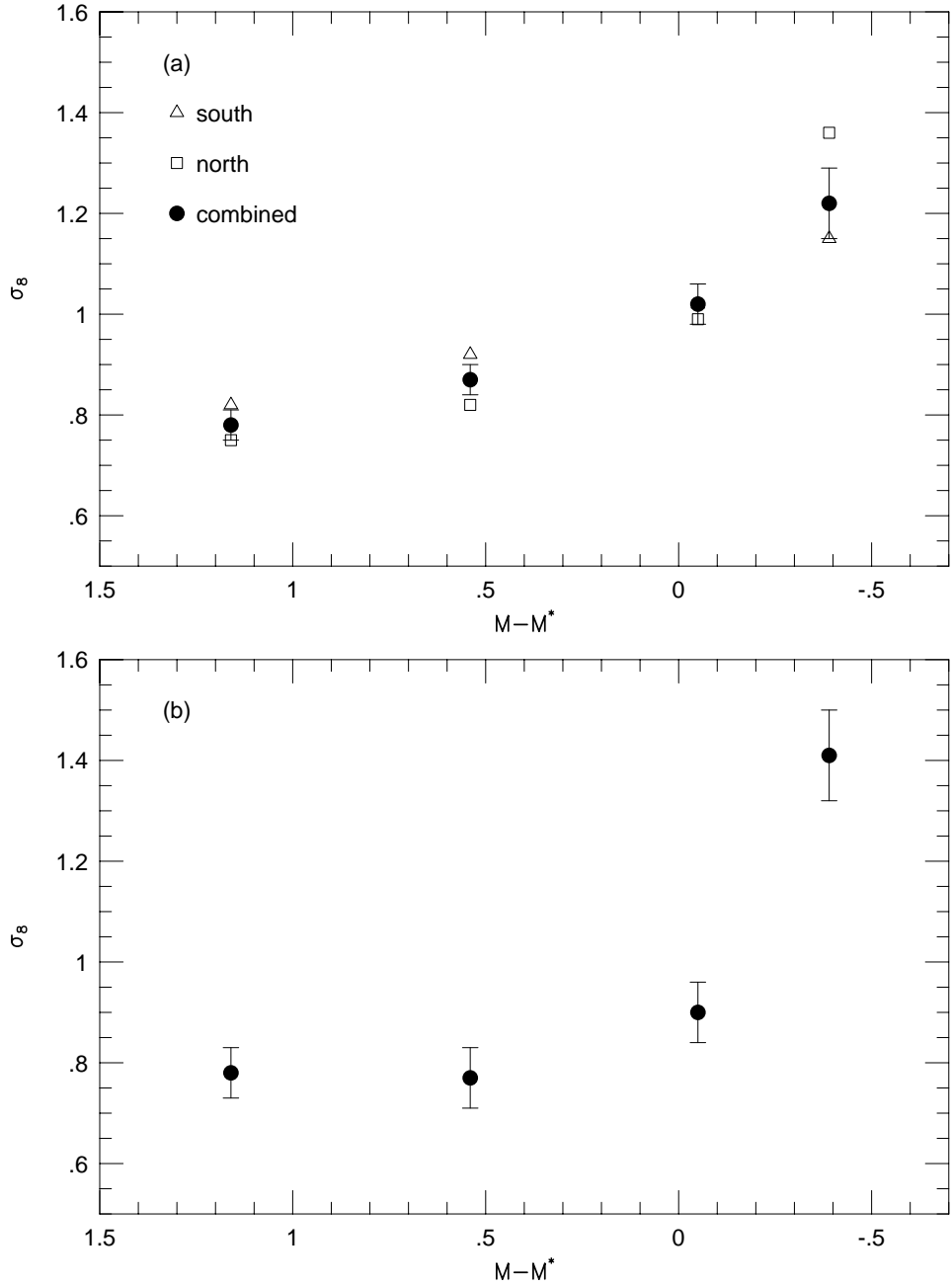}
\caption{Dependence of the variance as measured through $\sigma_8$ on
the galaxy luminosity using the correlation function measured in
redshift space (panel a) and real space (panel b).
In both panels the reference luminosity corresponds to the sample with $L$
$\sim$ $L^*$. The symbols in panel (a) represent galaxies of the SSRS2
south subsample (open triangles); SSRS2 north (open squares) and the
combined sample (solid circle).
The error bars were obtained by standard error propagation  combining
the bootstrap error estimates of the different samples, calculated at
each of the separations shown in the plot. In panel (b) we only show
the results obtained for the combined sample.}
\end{figure}
\clearpage

\begin{figure}
\vspace{185mm}
\includegraphics{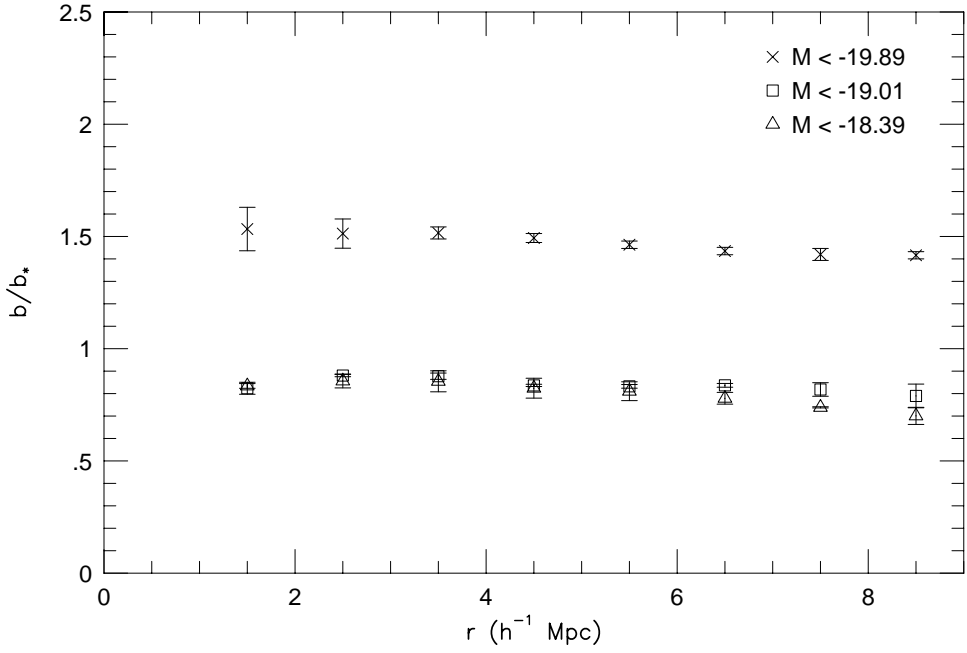}
\caption{Variation of the relative bias between galaxies of different
luminosities with scale, using the real space correlation function and
equation (15). In
this plot the reference luminosity corresponds to the sample with $L$
$\sim$ $L^*$. The symbols represent galaxies with $M \leq -19.89$
(crosses); $M \leq -19.01$ (squares) and $M \leq -18.39$ (triangles).
The error bars were obtained by standard error propagation  combining
the bootstrap error estimates of the different samples, calculated at
each of the separations shown in the plot.}
\end{figure}
\clearpage

\begin{figure}
\vspace{185mm}
\includegraphics{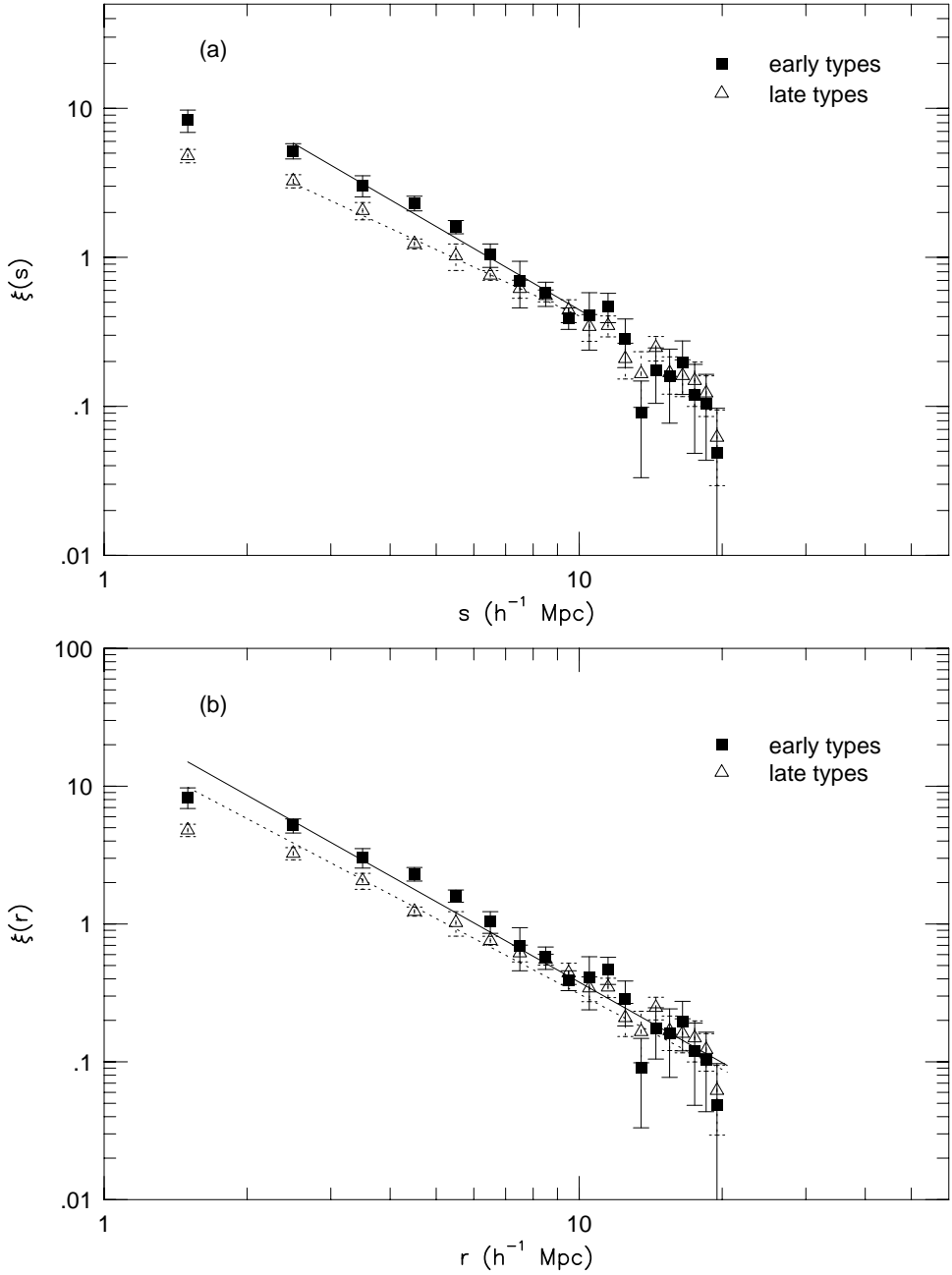}
\caption{The two-point correlation for the SSRS2 discriminating
morphological types calculated in redshift space (panel a). Solid
squares represent early type galaxies and open triangles late
types. The power-law fits are shown as solid and dotted lines
respectively for early and late types. In panel (b) we show the points
measured in redshift space together with the power-law fits measured in real
space. The errors were estimated from the bootstrap resampling.}
\end{figure}
\clearpage

\begin{figure}
\vspace{185mm}
\includegraphics{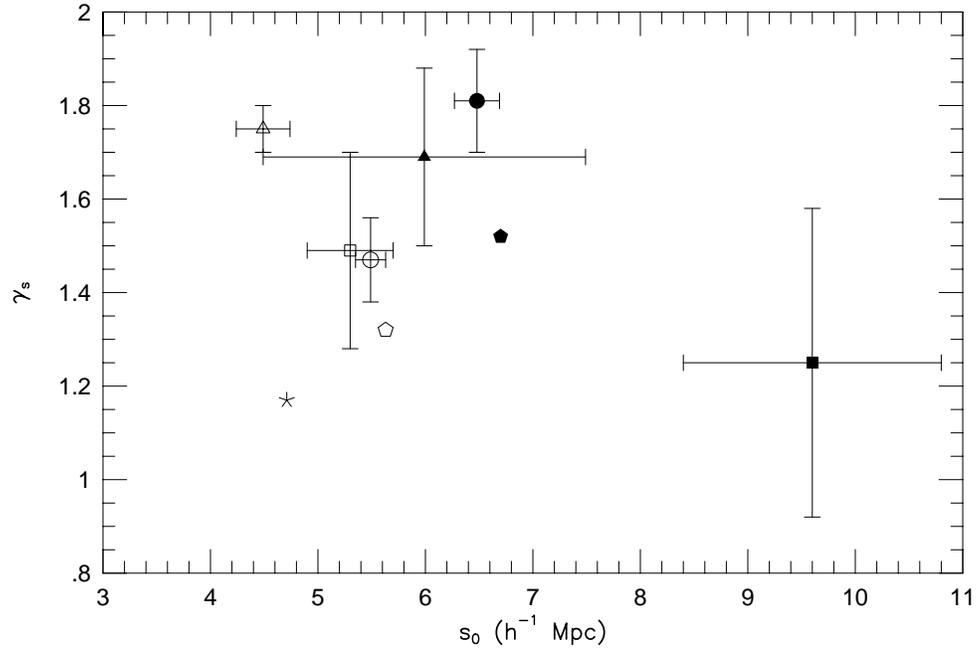}
\caption{Comparison between values of $s_0$ measured in this paper for
early types (solid circle) and late types (open circle), with measures
of other works, presented in Table 5. These are the Stromlo-APM,
represented as squares (solid for early types, open for late types),
SSRS1 (solid and open triangles for early and late types respectively)
and the ORS (early types as solid pentagon, Sa/Sb as open pentagon and
Sc/Sd as a star). The error bars for the SSRS2 sample represent the
statistical uncertainty obtained from the power-law fits.} 
\end{figure}
\clearpage

\begin{figure}
\vspace{185mm}
\includegraphics{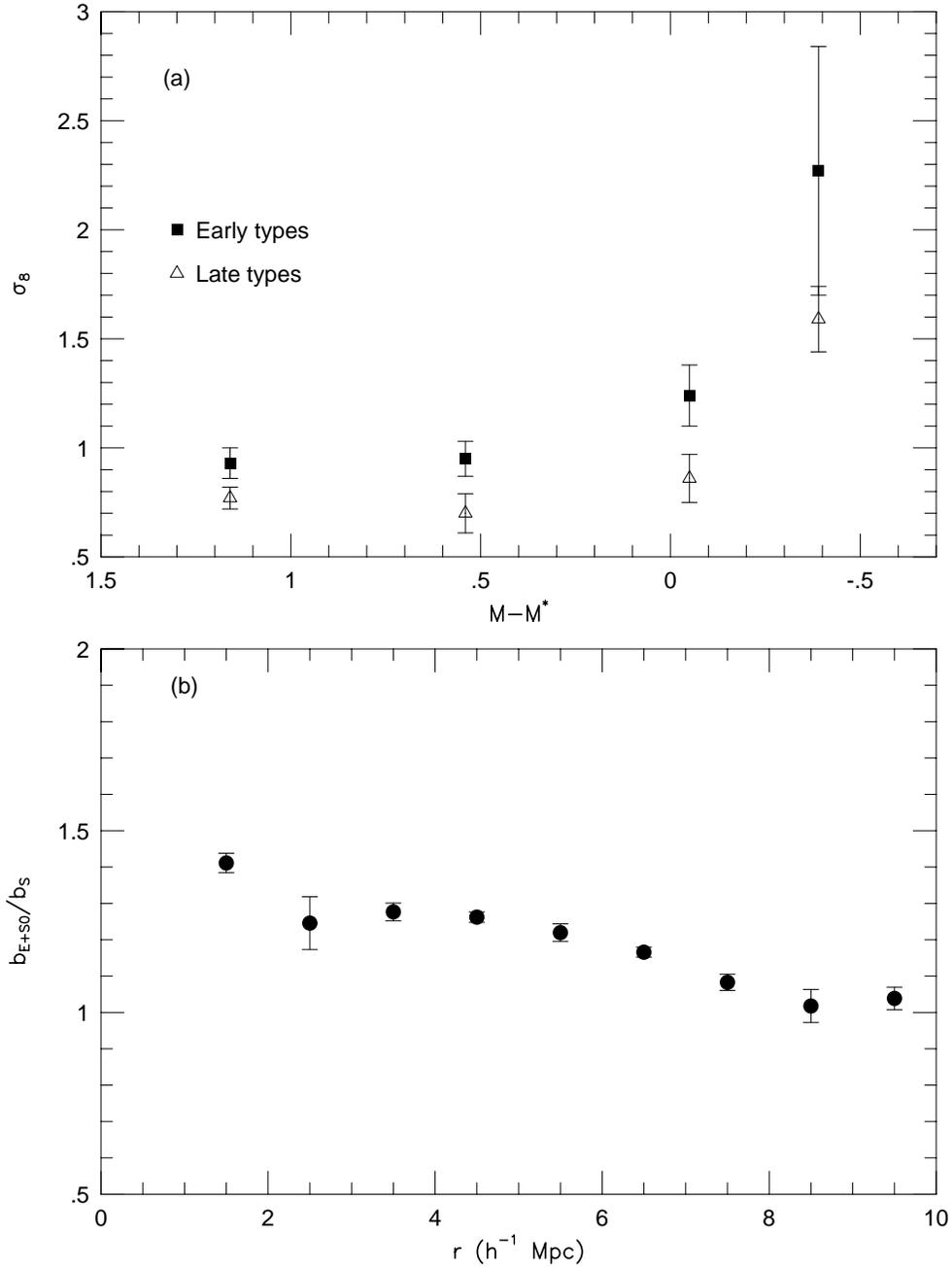}
\caption{Biasing measures for galaxies discriminated by
morphologies. Panel (a) shows the variance for different luminosity
thresholds while panel (b) shows the relative bias between early and
late types as a function of scale, using equation (16). In both panels
these measures are derived from the real space correlation function. 
The error bars  were derived through standard error propagation, using
in panel (a) the estimated errors of the parameters
obtained calculating the power-law fits, while in panel (b) they were
obtained combining the bootstrap error estimates of both 
samples, calculated at each of the separations shown in the plot. 
} 
\end{figure}
\clearpage

\begin{figure}
\vspace{185mm}
\includegraphics{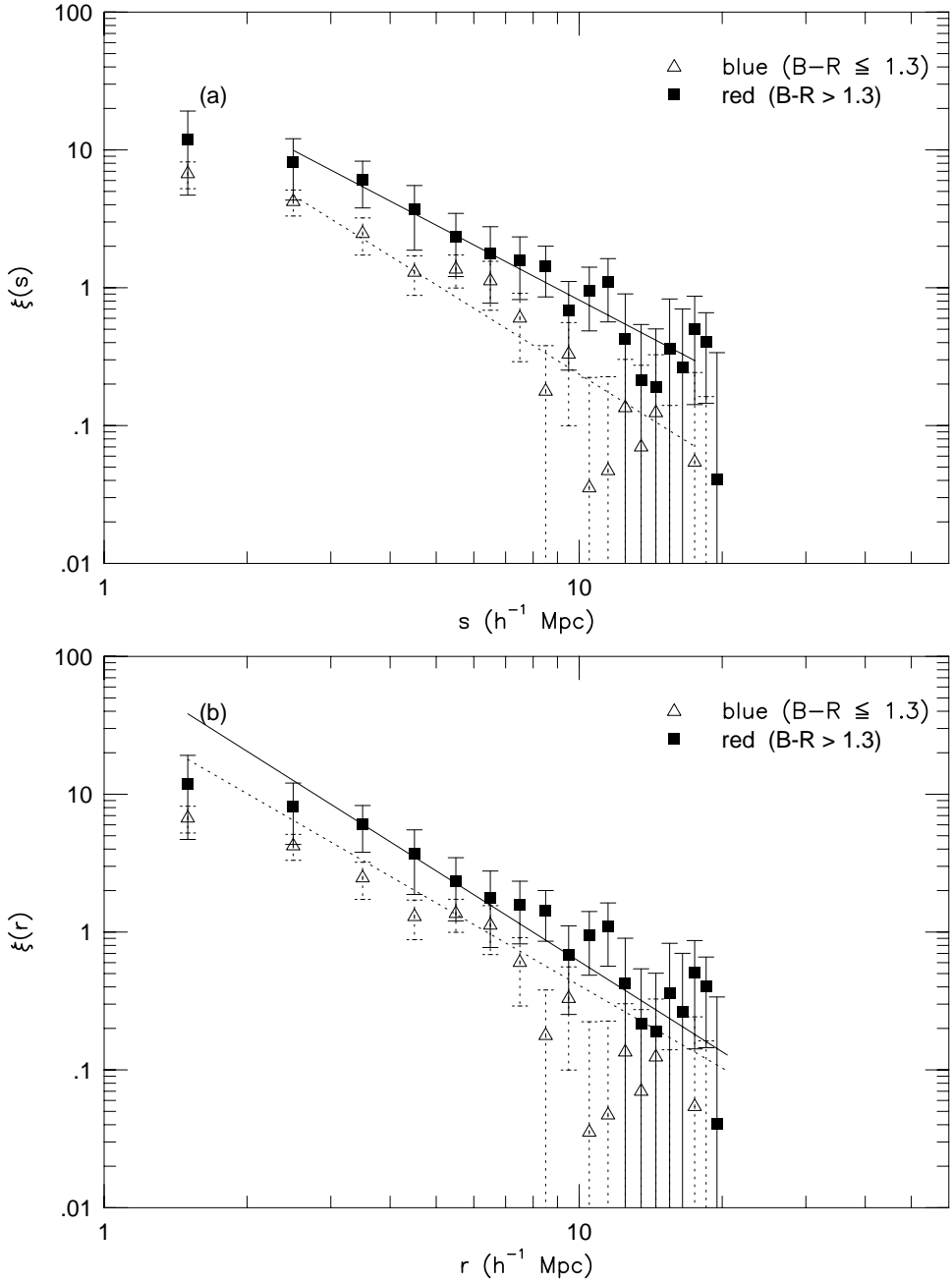}
\caption{Panel (a) shows the redshift space correlation for galaxies
discriminated now by colors. Table 8 shows the correlation parameters
for these fits. Panel (b) shows the real space correlation for galaxies
discriminated  by colors. The points show the redshift-space
correlation. Table 8 also shows the correlation parameters for these
fits. The error bars were obtained by bootstrap resampling.}
\end{figure}
\clearpage

\begin{figure}
\vspace{185mm}
\includegraphics{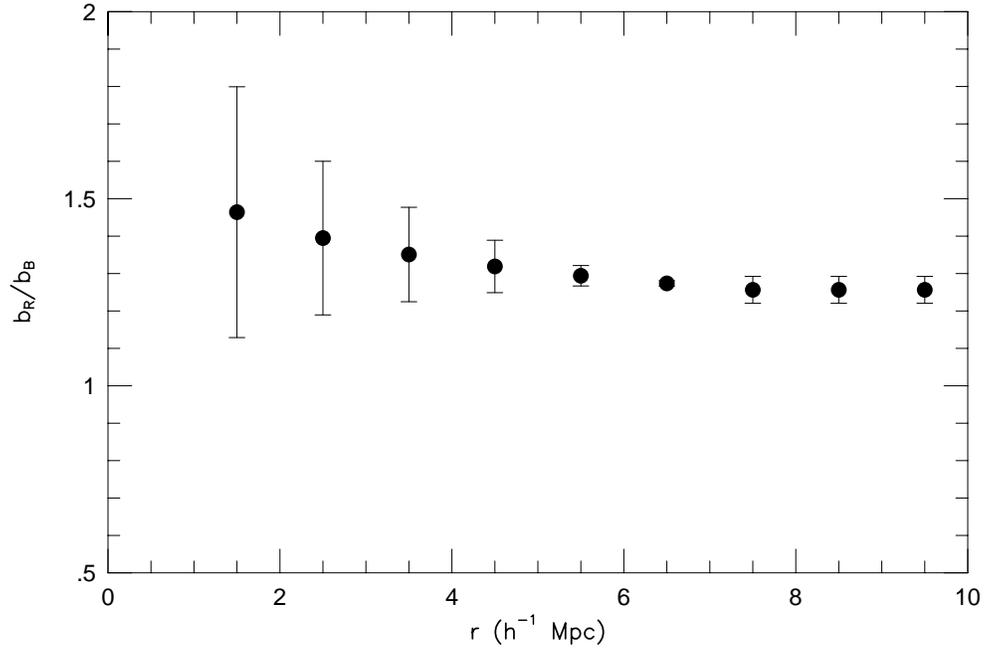}
\caption{Relative bias between galaxies with red and blue colors as a
function of scale, using the fits to $\xi(r)$. The error bars were
obtained combining the statistical errors obtained in the power-law
fits using standard error propagation.} 
\end{figure}
\clearpage

\begin{figure}
\vspace{185mm}
\includegraphics{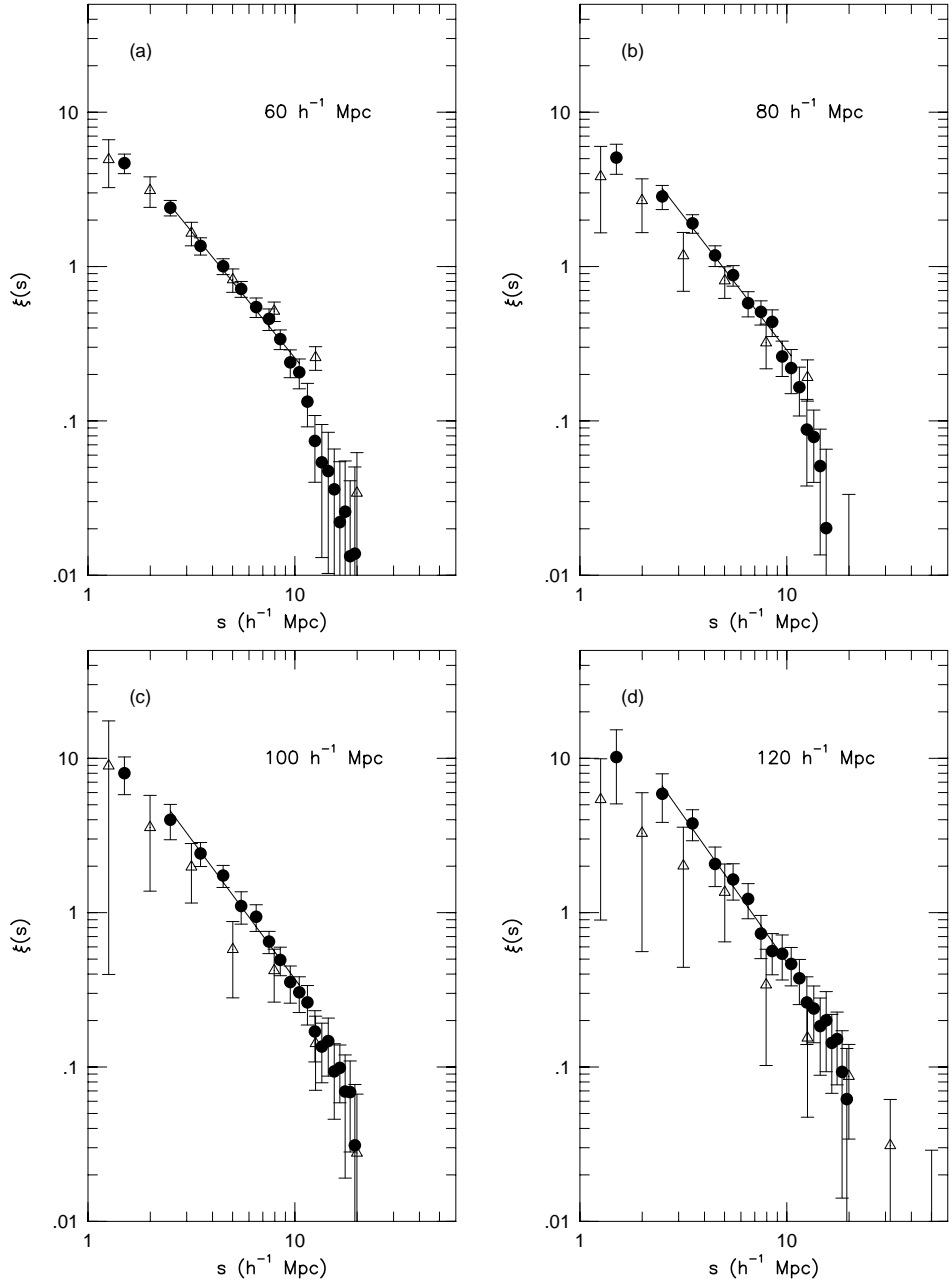}
\caption{Comparison between the redshift correlation of the
combined SSRS2 sample (solid circles) with that measured for $IRAS$
galaxies (open triangles), for the different volume limits indicated
in the panels. We also show the fits calculated for the SSRS2
sample. The error bars for both SSRS2 and $IRAS$ subsamples were
estimated using bootstrap resampling.}
\end{figure}
\clearpage

\begin{figure}
\vspace{185mm}
\includegraphics{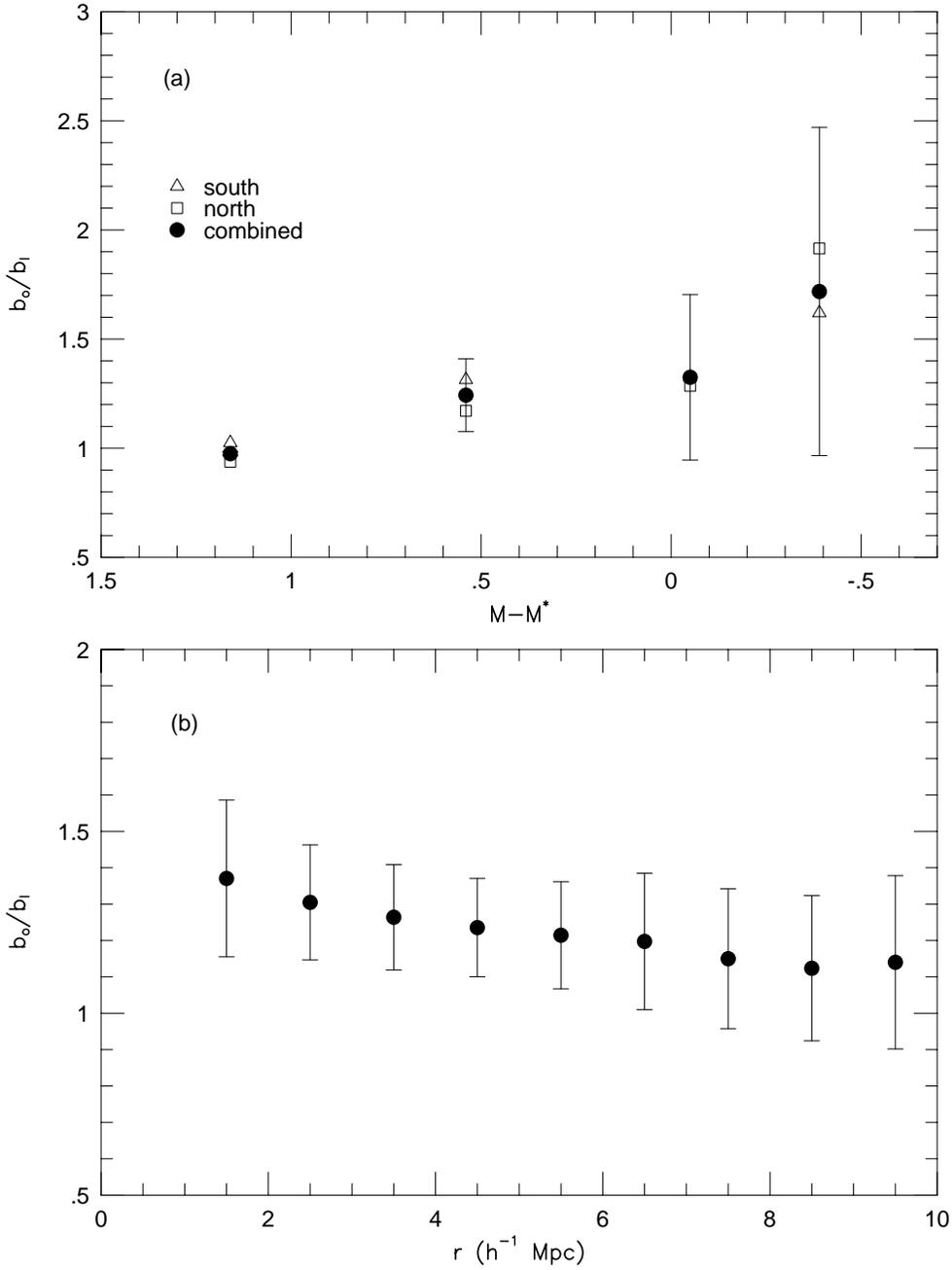}
\caption{Panel (a) shows the relative bias between optical and $IRAS$
galaxies as a function of luminosity, derived from the redshift space
correlation function. The symbols represent the relative bias obtained
considering the SSRS2 south (open triangles), SSRS2 north (open
squares) and the combined sample (solid circle). Panel (b) presents
the relative bias as a function of scale, but now using the real space
correlation function. The error bars were obtained
combining the $\sigma_8$ uncertainties for both samples (panel a), or
by combining the correlation function bootstrap error at each
separation in the case of panel (b). The large errors seen in panel
(b) reflect the larger uncertainty that is obtained when calculating
the correlation function in real space.}
\end{figure}
\clearpage

\end{document}